# Simple and Combination Parametric Resonances of an Electromagnetically Suspended Vehicle subject to Base Excitation

Jithu Paul[a,1], Karel N. van Dalen[a], Andrei B. Fărăgău[a], Rens J. van Leijden[a], Biagio Carboni[b], Andrei V. Metrikine[a]

[a] *Department of Engineering Structures, Faculty of CEG, TU Delft, NL*

[b] *Department of Structural and Geotechnical Engineering, Sapienza University of Rome, Italy*

This paper investigates the dynamic stability of an electromagnetically suspended vehicle, encountered in Hyperloop and Maglev systems, subject to purely sinusoidal base excitation caused by surface irregularities or vibration of the support induced by external noise. The narrow airgap between the vehicle and the support makes the system sensitive to the motion of the support, as small amplitudes of the latter create significant excitation. The vehicle is modelled as a three-degree-of-freedom model where the vehicle is suspended via two identical electromagnetic actuators from rigid supports that oscillate. The governing equations are derived using force and torque balances, incorporating nonlinear electromagnetic forces, and Kirchhoff's law for the electromagnets with PD control strategy on the airgap. The equations of motion are linearized around the steady state induced by the surface oscillation, yielding a system with time-periodic coefficients. We analytically explore both simple and combination parametric resonances using an extended Hill's method, and Floquet theory is used for numerical validation. The stability boundaries are obtained as ellipses in the PD control parameter space, and the influence of system parameters on these boundaries is characterized. For the simple parametric resonance, the ratio of the sizes of the two obtained ellipses is three to one, whereas for the combination parametric resonance, the ratio is fourteen to one. When all ellipses are simultaneously present, one of the ellipses associated with the combination parametric resonance is the largest. Moreover, we found that in all cases, the relative sizes of the ellipses are independent of the excitation frequency, when normalized by the local width of the stable domain. Additionally, the impact of using hybrid magnets in the supports—combining electromagnets with permanent magnets—on the parametric resonances is analysed, showing that they are equivalent to those of the electromagnet-only case; the ellipses only shift in accordance with an overall widening of the stable domain. Results reveal critical conditions under which each type of resonance dominates, offering key insights for safe design and operation of magnetically suspended vehicles.

## 1. Introduction

Magnetic levitation technologies are at the forefront of next-generation transport systems due to their potential for high-speed, frictionless travel. Among these, the Hyperloop concept—an electromagnetically suspended pod traveling through a near-vacuum tube—has attracted significant attention. Despite extensive advancements in vehicle design, control systems, and propulsion, the stability of these systems under dynamic excitations remains poorly explored. Recent textbooks provide comprehensive coverage of the practical aspects of Maglev train control [1,2].

For Maglev and Hyperloop designs using electro-magnetic suspension (EMS), the vehicle typically hovers just one or a few centimetres below a track. At such small gaps, even minor external perturbations—due to track irregularities, structural vibrations, or aerodynamic disturbances—can lead to dynamic instabilities such as various forms of parametric resonance. These phenomena are particularly important to understand as they directly affect ride safety, system integrity, and speed limitations [2].

The stability of vehicles moving along flexible guideways has been investigated from multiple perspectives [3]. Metrikine [4] demonstrated that anomalous Doppler waves can lead to dynamic instability at high speeds. Aeroelastic effects such as galloping, flutter, and vortex-induced vibrations are also known to influence stability, particularly in lightly damped, slender structures [5–8]. These phenomena have been extensively studied in the contexts of aircraft and railway vehicles.

[1] jithupaulv@gmail.com

In the domain of Maglev systems, several studies have analysed the effects of magnetic and electromagnetic force characteristics, and structural flexibility. Cai et al. [9] analysed the dynamic instability of electrodynamic Maglev systems by modelling three-degrees-of-freedom and five-degrees-of-freedom vehicles moving along a double L-shaped guideway, considering both steady magnetic forces and motion-dependent magnetic force coefficients. In another paper by the same authors [10], they explored the vehicle/guideway interactions in Maglev systems, focusing on how multiple cars and loads affect stability. The paper highlights the impact of vehicle/guideway coupling, compares concentrated and distributed loading, and evaluates ride comfort on single-span and double-span flexible guideways. A review on the dynamic stability of repulsive-force Maglev suspension systems is available in [11]. More complex systems that involve an interaction between different mechanisms can be found in the following literature. Wu et al. [12] examined suspension stability under the interaction of aeroelastic and electromagnetic mechanisms, while Wang et al. [13] and Zhang et al. [14] explored the destabilizing role of sensor and controller delays. Fărăgău et al. [15,16] highlighted the interplay between electromagnetic and wave-induced mechanisms and mapped the resulting regions of dynamic stability and occurrence of limit cycles.

A severe dynamic instability can be expected when considering coupling effects between different degrees of freedom of electromagnetically suspended vehicles, especially in the presence of movements of the suspension caused by either external excitation or infrastructure flexibility. Detailed investigations have been carried out in studies involving only a translation degree of freedom for the vehicle suspended from a single oscillating point: one considering the interaction of the electromagnetic suspension system with aeroelastic forces [17] and another considering the interaction of the suspension system with vibrations of the flexible periodic infrastructure [18]. However, real Maglev and Hyperloop vehicles are suspended at multiple points, so rotational motion and coupling between rotational and translational motions can naturally occur. The presence of a second suspension point is not a trivial extension of the single-suspension model; it introduces an additional instability type, namely combination parametric resonance, which is absent in the translation-only configuration. As shown in this paper, this additional instability can in some cases be the dominant one. A rigorous study on combination resonance for purely mechanical systems was conducted by Wanda Szemplińska-Stupnicka [19], who extended the harmonic balance method for parametrically excited systems. Numerous papers have emphasized the critical role of combination resonance in the dynamics of complex structures [20–26].

However, simple and combination parametric resonances have not yet been thoroughly investigated for Maglev and Hyperloop systems, although their relevance has been noted by [9–11] and others. In this study, we examine the significance of simple and combination parametric resonances as a function of vehicle speed. The aim is to gain understanding of which parameters are critical for such instabilities. This benchmark study provides valuable insights that can serve as a foundation for developing more sophisticated models representing more realistic scenarios.

Although the focus of the current paper is on transportation applications of Maglev and Hyperloop systems—particularly electromagnetic suspension (EMS)—the technology has also been widely utilized in other domains. Magnetic bearings [27–29], for instance, eliminate the need for lubrication systems by enabling contactless operation between the rotor and stator. Similarly, non-contact electromagnetic control [30,31] is used in the deployment of offshore wind turbine structures. These examples highlight the broad relevance and applicability of the findings presented in this study.

The paper is structured as follows. Section 2 presents the problem statement. Section 3 discusses the steady-state conditions and the linearized equations. Section 4 explores the types of parametric resonances possible in the system and analytically derives all the associated instability boundaries. Section 5 examines the effect of the hybrid magnet on these instability boundaries. Finally, Section 6 provides the conclusions.

## 2. Problem statement

In this paper we investigate the parametric resonances of the system shown in Fig. 1. This is an extension of the problem explored in [17], where we considered a single PD-controlled suspension. Here, we investigate the effect of a two-point suspension, which add rotational motion and enables combination parametric resonance, an additional instability type absent in the single-suspension system studied in [17]. The results may be applicable to any magnetically levitated/suspended mass with at least 2 degrees of freedom that is actively controlled and subject to external excitations—such as a Maglev train, an Hyperloop vehicle, a magnetic bearing, or a magnetic pendulum used in offshore structure deployment. Specifically, we have chosen parameters corresponding to a scaled Hyperloop system [15], and the paper is developed accordingly.

The support from which the vehicle is suspended can undergo oscillations due to external noise or surface roughness leading to oscillations of the support of the moving vehicle. In systems like Maglev or Hyperloop using EMS, the gap between the support and the vehicle is typically only a one or a few centimetres. The comparable dimensions of undisturbed airgap and irregularity make the study of small-amplitude irregularity particularly significant; the irregularity can induce parametric resonance. Understanding the instabilities of the time-periodic steady state (which is induced by the external excitation) is crucial for designing supporting structures and determining speed limits for the vehicle.

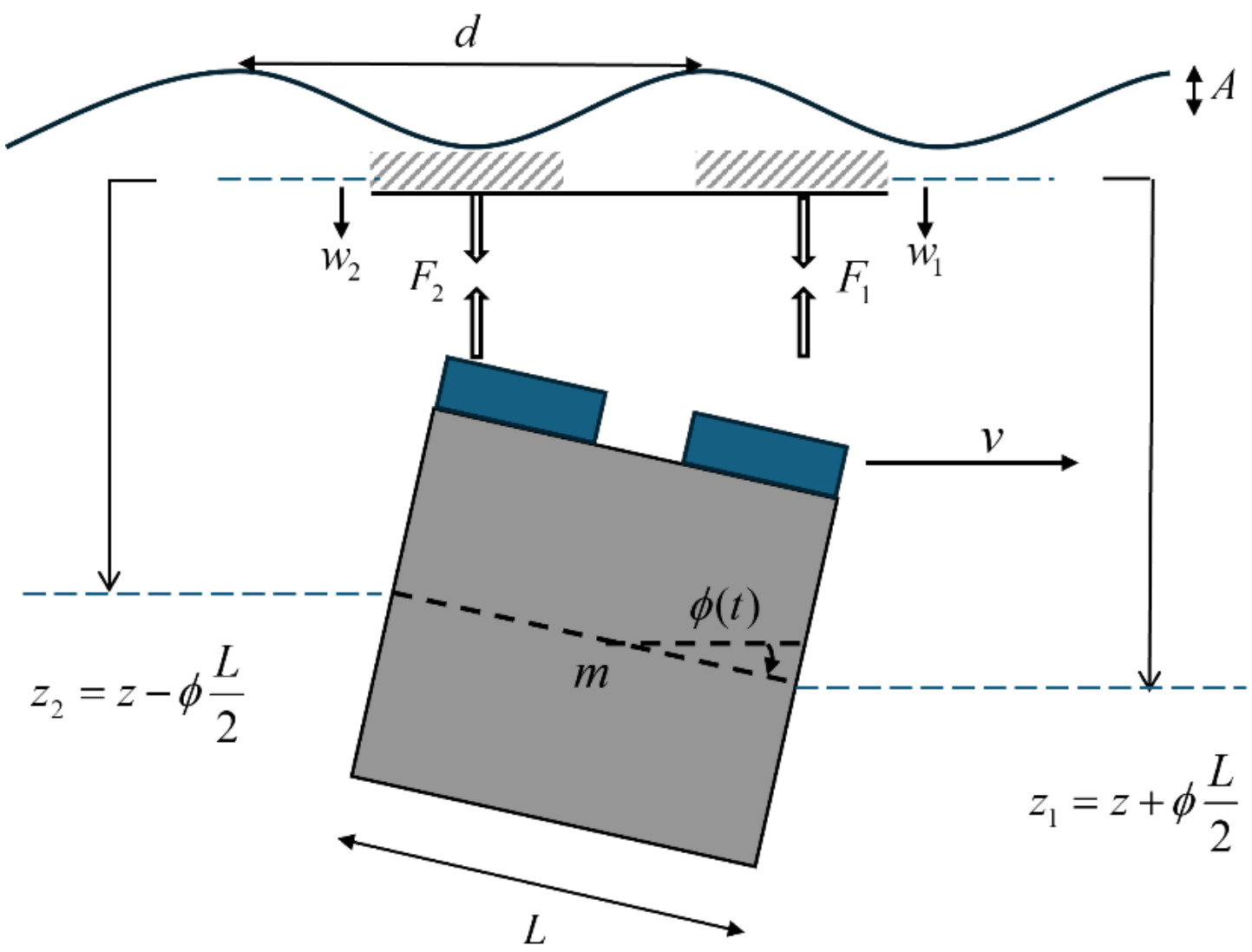

Fig. 1. Model system.

As shown in Fig. 1, we consider a model of an electromagnetically suspended vehicle hanging from two points (subscripts of variables labelled as 1 and 2), with mass $m$, rotational moment of inertia $J$, and length $L$. The vehicle moves with speed $v$. The rigid support has an ideal wavy surface characterized by wavelength $d$, which induces an (imposed) oscillation to the moving vehicle's suspensions characterized by frequency $\Omega$, phase shift $\theta$, and amplitude $A$. Here, $\Omega = 2\pi/T = 2\pi v/d$ (as $T = d/v$) is the temporal excitation frequency experienced by the moving vehicle, and $\theta$ is the phase difference between the excitations at the two suspension points separated by the distance $L$.

The phase difference stems from the delay (of the experienced wavy profile) between the two supports: $\cos\left(\Omega(t-\Delta t)\right)=\cos\left(\Omega t-\theta\right)$, with $\Delta t = L/v$, so that $\theta=\Omega\Delta t=2\pi L/d$ .

The vehicle is subject to two similar electromagnetic forces acting at points 1 and 2, and a gravitational force acting at its centre of mass. Due to the gap $L$ between the electromagnetic actuators, a torque is also generated. The force and torque balances yield the following equations of motion (EOMs), with $z$ and $\phi$ being the translation and the rotation about the centre of gravity, respectively:

$$m\ddot{z}=-C\left(\frac{I_1^2}{\Delta_1^2}+\frac{I_2^2}{\Delta_2^2}\right)+mg \tag{1}$$

$$J\ddot{\phi}=-C\left(\frac{I_1^2}{\Delta_1^2}-\frac{I_2^2}{\Delta_2^2}\right)\frac{L}{2} \tag{2}$$

In Eqns. (1) and (2), $F_i=CI_i^2/\Delta_i^2$, $i=1,2$, represents the nonlinear electromagnetic force, where $C$ is the electromagnetic constant [2], and the overdot represents differentiation with respect to time. This force model is an idealized approximation conventionally used for EMS systems. Effects such as magnetic saturation and other large-displacement nonlinearities are outside the scope of the present study; however, they do not affect the linear-stability boundaries of the steady state considered here. Assuming small rotations of the vehicle about its centre of gravity, i.e., $|\phi|<<1$, the position of the vehicle supports are given by $z_{1,2}=z\pm\phi L/2$. The gaps $\Delta_{1,2}$ between the rigid oscillating supports (oscillation is perceived by the vehicle moving along the wavy pattern) and the vehicle is given by:

$$\Delta_{1,2}=z_{1,2}-w_{1,2};\ w_1=A\cos\left(\Omega t\right);\ w_2=A\cos\left(\Omega t-\theta\right) \tag{3}$$

The currents in Eqns. (1)-(2), for the two electromagnets are controlled by identical PD controllers. The control equations, derived from Kirchhoff's law [2], are as follows:

$$\dot{I}_{1,2}+\frac{\Delta_{1,2}}{2C}\left(R-2C\frac{\dot{\Delta}_{1,2}}{\Delta_{1,2}^2}\right)I_{1,2}=\frac{\Delta_{1,2}}{2C}U_{1,2} \tag{4}$$

In Eq. (4), the voltages $U_{1,2}$ consist of steady-state components and perturbation-induced components. The latter arise when the gaps $\Delta_{1,2}$ and their rates $\dot{\Delta}_{1,2}$ deviate from the steady-state counterparts, which activates the PD controller:

$$U_{1,2}=U_{\mathrm{ss1,\,ss2}}+K_{\mathrm{p}}\left(\Delta_{1,2}-\Delta_{\mathrm{ss1,\,ss2}}\right)+K_{\mathrm{d}}\left(\dot{\Delta}_{1,2}-\dot{\Delta}_{\mathrm{ss1,\,ss2}}\right) \tag{5}$$

In Eq. (5), $K_{\mathrm{p}}$ and $K_{\mathrm{d}}$ are the PD controller parameters, while $\Delta_{\mathrm{ss1,\,ss2}}$ represent the steady-state components of the gaps at suspensions 1 and 2, respectively. Clearly, in the steady state, the PD controller is not active.

We note that the model has three degrees of freedom: $z$ and $\phi$ both correspond to a second-order equation of motion and therefore represent one degree of freedom each, while $I_1$ and $I_2$ both correspond to a first-order electrical equation and are therefore counted as half a degree of freedom each.

## 3. Steady state and linearised equations

Eqns. (1), (2) and (4) are nonlinear; for the stability analysis, we linearise the EOMs about the steady-state response. In Eqns. (1)-(4), at the steady state, there are 2 pairs of equations and three pairs of dependent variables $z_{\mathrm{ss1,\,ss2}}$, $I_{\mathrm{ss1,\,ss2}}$ and $U_{\mathrm{ss1,\,ss2}}$. Hence one pair of variables must be prescribed to close the definition of the steady state [17,18], and we assume the steady-state positions $z_{\mathrm{ss1,\,ss2}}$ to be constant and equal to $z_0$ (i.e., $\phi_{\mathrm{ss}} = 0$). At the steady state, Eqns. (1) and (2) yield the gaps $\Delta_{\mathrm{ss1,\,ss2}}$ and currents $I_{\mathrm{ss1,\,ss2}}$, and Eq. (4) gives the voltages $U_{\mathrm{ss1,\,ss2}}$ based on those:

$$\Delta_{\mathrm{ss1}} = z_{\mathrm{ss1}} - A\cos(\Omega t);\ \Delta_{\mathrm{ss2}} = z_{\mathrm{ss2}} - A\cos(\Omega t - \theta);\ I_{\mathrm{ss1,\,ss2}} = \sqrt{\frac{mg}{2C}}\Delta_{\mathrm{ss1,\,ss2}};\ U_{\mathrm{ss1,ss2}} = RI_{\mathrm{ss1,\,ss2}} \tag{6}$$

To linearise Eqns. (1), (2), and (4), we introduce small perturbations (denoted by the subscript "tr") around the steady state as,

$$z_{1,2}(t) = z_{\mathrm{ss1,\,ss2}} + \Delta_{\mathrm{tr1,\,tr2}}(t);\ I_{1,2}(t) = I_{\mathrm{ss1,\,ss2}}(t) + I_{\mathrm{tr1,\,tr2}}(t) \tag{7}$$

leading to:

$$\frac{m}{2}\left(\ddot{\Delta}_{\mathrm{tr1}} + \ddot{\Delta}_{\mathrm{tr2}}\right) - \frac{2CI_{\mathrm{ss1}}^2}{\Delta_{\mathrm{ss1}}^3}\Delta_{\mathrm{tr1}} + \frac{2CI_{\mathrm{ss1}}}{\Delta_{\mathrm{ss1}}^2}I_{\mathrm{tr1}} - \frac{2CI_{\mathrm{ss2}}^2}{\Delta_{\mathrm{ss2}}^3}\Delta_{\mathrm{tr2}} + \frac{2CI_{\mathrm{ss2}}}{\Delta_{\mathrm{ss2}}^2}I_{\mathrm{tr2}} = 0 \tag{8}$$

$$\frac{2J}{L^2}\left(\ddot{\Delta}_{\mathrm{tr1}} - \ddot{\Delta}_{\mathrm{tr2}}\right) - \frac{2CI_{\mathrm{ss1}}^2}{\Delta_{\mathrm{ss1}}^3}\Delta_{\mathrm{tr1}} + \frac{2CI_{\mathrm{ss1}}}{\Delta_{\mathrm{ss1}}^2}I_{\mathrm{tr1}} + \frac{2CI_{\mathrm{ss2}}^2}{\Delta_{\mathrm{ss2}}^3}\Delta_{\mathrm{tr2}} - \frac{2CI_{\mathrm{ss2}}}{\Delta_{\mathrm{ss2}}^2}I_{\mathrm{tr2}} = 0 \tag{9}$$

$$\dot{I}_{\mathrm{tr1,\,tr2}} + \frac{R\Delta_{\mathrm{ss1,\,ss2}}^2 - 2C\dot{\Delta}_{\mathrm{ss1,\,ss2}}}{2C\Delta_{\mathrm{ss1,\,ss2}}}I_{\mathrm{tr1,\,tr2}} - \left(\frac{K_{\mathrm{p}}}{2C}\Delta_{\mathrm{ss1,\,ss2}} - \frac{\dot{\Delta}_{\mathrm{ss1,\,ss2}}I_{\mathrm{ss1,\,ss2}}}{\Delta_{\mathrm{ss1,\,ss2}}^2}\right)\Delta_{\mathrm{tr1,\,tr2}} - \left(\frac{K_{\mathrm{d}}}{2C} + \frac{I_{\mathrm{ss1,\,ss2}}}{\Delta_{\mathrm{ss1,\,ss2}}^2}\right)\Delta_{\mathrm{ss1,\,ss2}}\dot{\Delta}_{\mathrm{tr1,\,tr2}} = 0 \tag{10}$$

Eliminating $I_{\mathrm{tr1}}$ and $I_{\mathrm{tr2}}$ from Eqns. (8)-(9) by using Eqn. (10) and substituting $\Delta_{\mathrm{tr}} = \left(\Delta_{\mathrm{tr1}} + \Delta_{\mathrm{tr2}}\right)/2;\ \phi_{\mathrm{tr}} = \left(\Delta_{\mathrm{tr1}} - \Delta_{\mathrm{tr2}}\right)/L$ results in the following simplified EOMs:

$$\begin{aligned}&\left(-8K_{\mathrm{p}}L\sqrt{Cgm} + 4\sqrt{2}gLmR\right)\Delta_{\mathrm{tr}}(t) - 8K_{\mathrm{d}}L\sqrt{Cgm}\dot{\Delta}_{\mathrm{tr}}(t) - \\ &-\left(\sqrt{2}LmR\Delta_{\mathrm{ss1}}(t) + \sqrt{2}LmR\Delta_{\mathrm{ss2}}(t)\right)\ddot{\Delta}_{\mathrm{tr}}(t) - \left(2\sqrt{2}JR\Delta_{\mathrm{ss1}}(t) - 2\sqrt{2}JR\Delta_{\mathrm{ss2}}(t)\right)\ddot{\phi}_{\mathrm{tr}}(t) - 4\sqrt{2}CLm\dddot{\Delta}_{\mathrm{tr}}(t) = 0\end{aligned} \tag{11}$$

$$\begin{aligned}&\left(-4K_{\mathrm{p}}L^2\sqrt{Cgm} + 2\sqrt{2}gL^2mR\right)\phi_{\mathrm{tr}}(t) - 4K_{\mathrm{d}}L^2\sqrt{Cgm}\dot{\phi}_{\mathrm{tr}}(t) - \\ &-\left(\sqrt{2}LmR\Delta_{\mathrm{ss1}}(t) - \sqrt{2}LmR\Delta_{\mathrm{ss2}}(t)\right)\ddot{\Delta}_{\mathrm{tr}}(t) - \left(2\sqrt{2}JR\Delta_{\mathrm{ss1}}(t) + 2\sqrt{2}JR\Delta_{\mathrm{ss2}}(t)\right)\ddot{\phi}_{\mathrm{tr}}(t) - 8\sqrt{2}CJ\dddot{\phi}_{\mathrm{tr}}(t) = 0\end{aligned} \tag{12}$$

where

$$\begin{aligned}&\Delta_{\mathrm{ss1}}(t) = z_0 - A\cos(\Omega t)\\ &\Delta_{\mathrm{ss2}}(t) = z_0 - A\cos(\Omega t - \theta) = z_0 - \left(P\cos(\Omega t) + Q\sin(\Omega t)\right);\\ &P = A\cos\theta;\ Q = A\sin\theta\end{aligned} \tag{13}$$

In the next section, we will explore parametric resonances. Before that, let us analyze the possible natural frequencies of the system shown in Fig. 1, based on the linearized system without parametric excitation, i.e., for $A=0$, and the same equilibrium position $z_0$. In this case, the coefficients are time independent, and the natural frequencies are obtained from the regular eigenvalue problem associated with the corresponding constant-coefficient linearized system. We will have two sets of three eigenvalues, i.e., six eigenvalues in total, as the system consists of two uncoupled subsystems, with 1.5 degrees of freedom each. At the stability boundary where parametric resonances are expected [17], each set of eigenvalues contains one real eigenvalue and a pair of complex-conjugate purely imaginary eigenvalues. Therefore, we expect the system to have two natural frequencies, defined as $\omega_1$ and $\omega_2$. At the stability boundary, the natural frequency associated with the translational/vertical vibration can be obtained through a simple substitution $\Delta_{\mathrm{tr}}=B\cos(\omega_1 t);\ \phi_{\mathrm{tr}}=0;\ A=0$, where $B$ is an undetermined constant, in Eqs. (11), and (12). Similarly, the natural frequency associated with the rotational vibration can be determined using the substitution $\Delta_{\mathrm{tr}}=0;\ \phi_{\mathrm{tr}}=B\cos(\omega_2 t);\ A=0$ in Eqs. (11), and (12). The obtained natural frequencies read as follows (see also Fig. 2):

$$\omega_2=\sqrt{3}\omega_1;\ \omega_1=\sqrt{K_{\mathrm{d}}}\left(\frac{2g}{mC}\right)^{1/4} \tag{14}$$

The stability analysis for the unexcited system can thus be performed using standard eigenvalue analysis, and the stability boundaries are shown as vertical and inclined black lines in Figs. 3–5 (the equilibrium is stable in between); the right stability boundary is related to an oscillatory, flutter-type instability (supercritical Hopf bifurcation, as addressed above), while a divergence instability emerges at the left boundary.

## 4. Parametric resonance

It is evident that Eqs. (11) and (12) form a homogeneous linear system with time-periodic coefficients; therefore, the stability of the free vibration must be examined for possible parametric resonances. For systems that can only oscillate vertically, only simple parametric resonance can exist. However, for systems that can rotate too, and where the vertical translation and the rotation are coupled, combination resonance may also occur.

In the system shown in Fig. 1, both vertical and angular oscillations of the vehicle are possible, resulting in two natural frequencies. Along the right stability boundary (Fig. 3), where the parametric resonance zones appear, the frequencies are denoted as $\omega_1$ and $\omega_2$ (Eq. (14)). For $\theta=0$, the two simplified EOMs, Eqns. (11) and (12), decouple completely. That means, the coefficient of $\ddot{\phi}_{\mathrm{tr}}(t)$ in Eq. (11), $2\sqrt{2}JR\Delta_{\mathrm{ss1}}(t)-2\sqrt{2}JR\Delta_{\mathrm{ss2}}(t)=0$ and the coefficient of $\ddot{\Delta}_{\mathrm{tr}}$ in Eq. (12) is zero too, which means that combination parametric resonance is excluded; the only parametric forcing terms in the EOMs for Eq. (11) and Eq. (12) are related solely to $\Delta_{\mathrm{tr}}$ and $\phi_{\mathrm{tr}}$, respectively. Thus, only simple parametric resonance is possible.

For $\theta=\pi$, the EOMs are coupled, but in Eq. (11) the parametric forcing term depends solely on the variable $\phi_{\mathrm{tr}}$ and in Eq. (12) it solely depends on the variable $\Delta_{\mathrm{tr}}$. This configuration gives rise to exclusively combination parametric resonances.

For intermediate values of $\theta$, both EOMs contain parametric forcing terms involving both $\Delta_{\mathrm{tr}}$ and $\phi_{\mathrm{tr}}$; hence, both simple and combination parametric resonances are expected.

Fig. 2 shows the natural frequencies (defined at the right stability boundary) and their sum and difference as functions of $K_\mathrm{d}$. Simple parametric resonance ($T$ and $2T$, where $T = 2\pi/\Omega$) is expected when $\omega_1 = \Omega/2,\ \Omega;\ \omega_2 = \Omega/2,\ \Omega$, and combination parametric resonance occurs when $\omega_1 + \omega_2 = \Omega/2,\ \Omega;\ \omega_2 - \omega_1 = \Omega/2,\ \Omega$. Although the two dashed lines ($\Omega,\ \Omega/2$) intersect the curves in Fig. 2 at eight points, only four of these parametric resonances are actually observed in both the analytical and numerical Floquet analyses discussed in this paper: $\omega_1 = \Omega/2$ and $\omega_2 = \Omega/2$ for simple resonance, and $\omega_1 + \omega_2 = \Omega$ and $\omega_2 - \omega_1 = \Omega$ (see Eqns. (29)-(30)) for combination resonance. Other, higher-order resonances are not captured by the present first-order analytical approximation and were not observed in the numerical Floquet analysis either; therefore, they are not considered further.

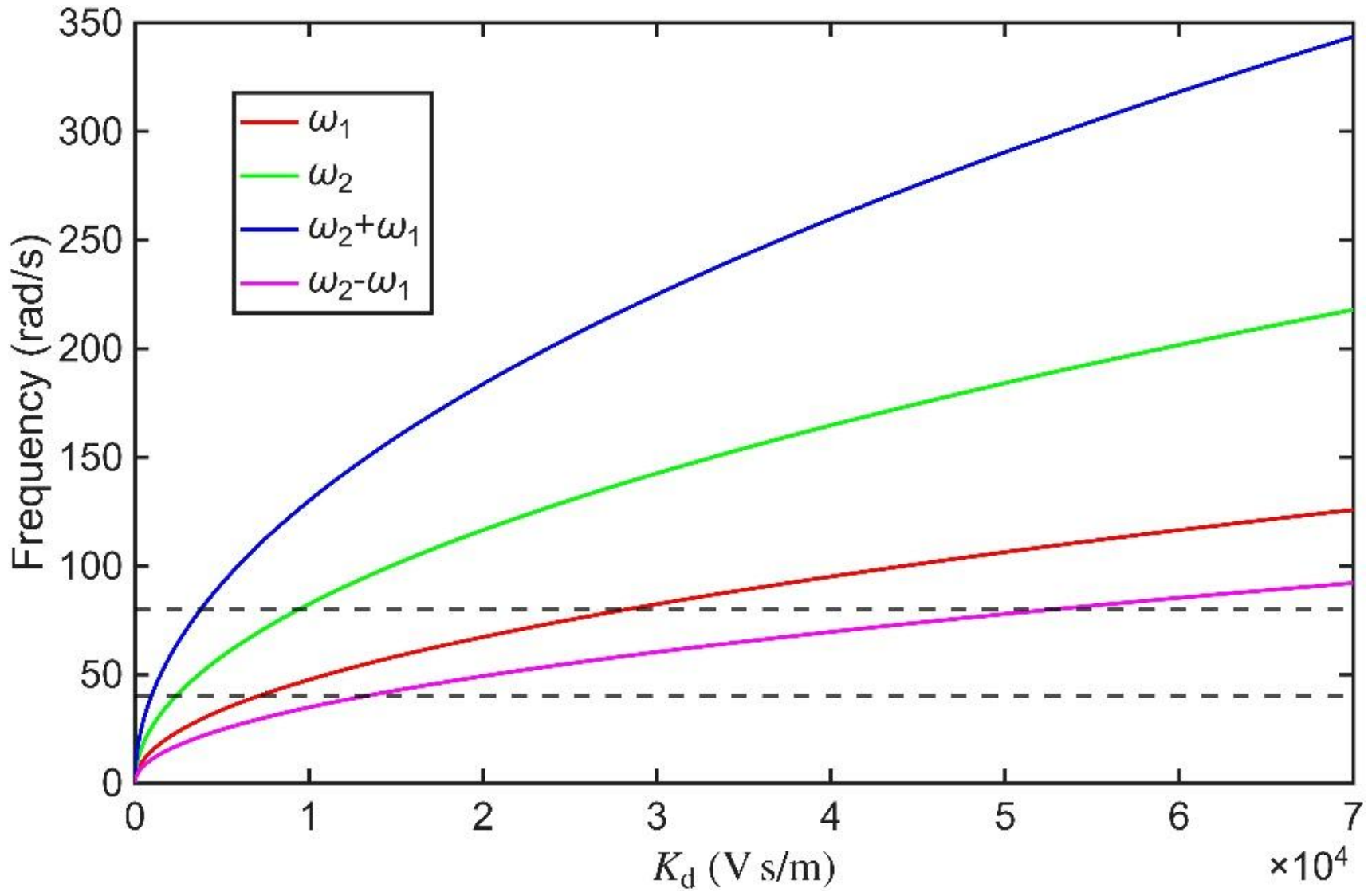

Fig. 2. Possible simple and combination parametric resonance frequencies. Here, $\omega_1$ and $\omega_2$ are translational and rotational natural frequencies. Here, $C = 0.05\left(\mathrm{Nm^2/A^2}\right)$, $z_0 = 0.015\left(\mathrm{m}\right)$, $m = 7650\left(\mathrm{kg}\right)$, $R = 9.71(\mathrm{Ohm})$, $A = 0.0142\left(\mathrm{m}\right)$, $\Omega = 80\left(\mathrm{rad/s}\right)$.

### 4.1. Simple parametric resonance

In this section, we derive the stability boundary of the system related to simple parametric resonance. For simple parametric resonance, as mentioned in the previous section we have the conditions $\omega_1 = \Omega/2;\ \omega_2 = \Omega/2$. For the first stability boundary, related to rotational degree of freedom (denoted as a), we assume the solution has the following form [19]:

$$
\begin{aligned}
&\Delta_\mathrm{tr}\left(t\right) = 0 \\
&\phi_\mathrm{tr}\left(t\right) = b_0 \cos\left(\omega_2 t\right) + b_1 \sin\left(\omega_2 t\right)
\end{aligned}
\tag{15}
$$

Substituting Eq. (15) into Eq. (12) and performing trigonometric reduction, while retaining terms proportional to fundamental harmonics, as assumed in Eq. (15), yields:

$$
S_1 \sin\left(\omega_2 t\right) + C_1 \cos\left(\omega_2 t\right) = 0 \tag{16}
$$

where $S_1$ and $C_1$ are given as:

$$
\begin{aligned}
&-4b_0 K_\mathrm{p} L^2 \sqrt{Cgm} + 2\sqrt{2} b_0 g L^2 m R - 4 b_1 K_\mathrm{d} L^2 \sqrt{Cgm}\,\omega_2 - \sqrt{2} A b_0 J R \omega_2^2 - \\
&-\sqrt{2} b_0 J P R \omega_2^2 - \sqrt{2} b_1 J Q R \omega_2^2 + 4\sqrt{2} b_0 J R z_0 \omega_2^2 + 8\sqrt{2} b_1 C J \omega_2^3 = 0
\end{aligned}
\tag{17}
$$

$$
\begin{aligned}
&-4b_1K_\mathrm{p}L^2\sqrt{Cgm}+2\sqrt{2}b_1gL^2mR+4b_0K_\mathrm{d}L^2\sqrt{Cgm}\omega_2+\sqrt{2}Ab_1JR\omega_2^2+\\
&+\sqrt{2}b_1JPR\omega_2^2-\sqrt{2}b_0JQR\omega_2^2+4\sqrt{2}b_1JRz_0\omega_2^2-8\sqrt{2}b_0CJ\omega_2^3=0
\end{aligned}
\tag{18}
$$

Extracting the truncated Hill's matrix [32] from Eqs. (17)-(18) and setting its determinant to zero gives the stability boundary in the form of an ellipse in the PD control parameter space. The centre of the ellipse is located at $\left(h_{1,\mathrm{a}},h_{2,\mathrm{a}}\right)$, and it has major axis $k_{1,\mathrm{a}}$ and minor axis $k_{2,\mathrm{a}}$ ($\omega_2$ has been replaced by $\Omega/2$):

$$
\frac{\left(K_\mathrm{p}-h_{1,\mathrm{a}}\right)^2}{k_{1,\mathrm{a}}^2}+\frac{\left(K_\mathrm{d}-h_{2,\mathrm{a}}\right)^2}{k_{2,\mathrm{a}}^2}=1
\tag{19}
$$

$$
\begin{aligned}
h_{1,\mathrm{a}}&=\frac{mR\left(24g+z_0\Omega^2\right)}{24\sqrt{2}\sqrt{Cgm}}\\
h_{2,\mathrm{a}}&=\frac{Cm\Omega^2}{12\sqrt{2}\sqrt{Cgm}}\\
k_{1,\mathrm{a}}&=\frac{\Omega}{2}k_{2,\mathrm{a}}\\
k_{2,\mathrm{a}}&=\sqrt{\frac{A^2mR^2\Omega^2\left(1+\cos\theta\right)}{2304Cg}}
\end{aligned}
\tag{20}
$$

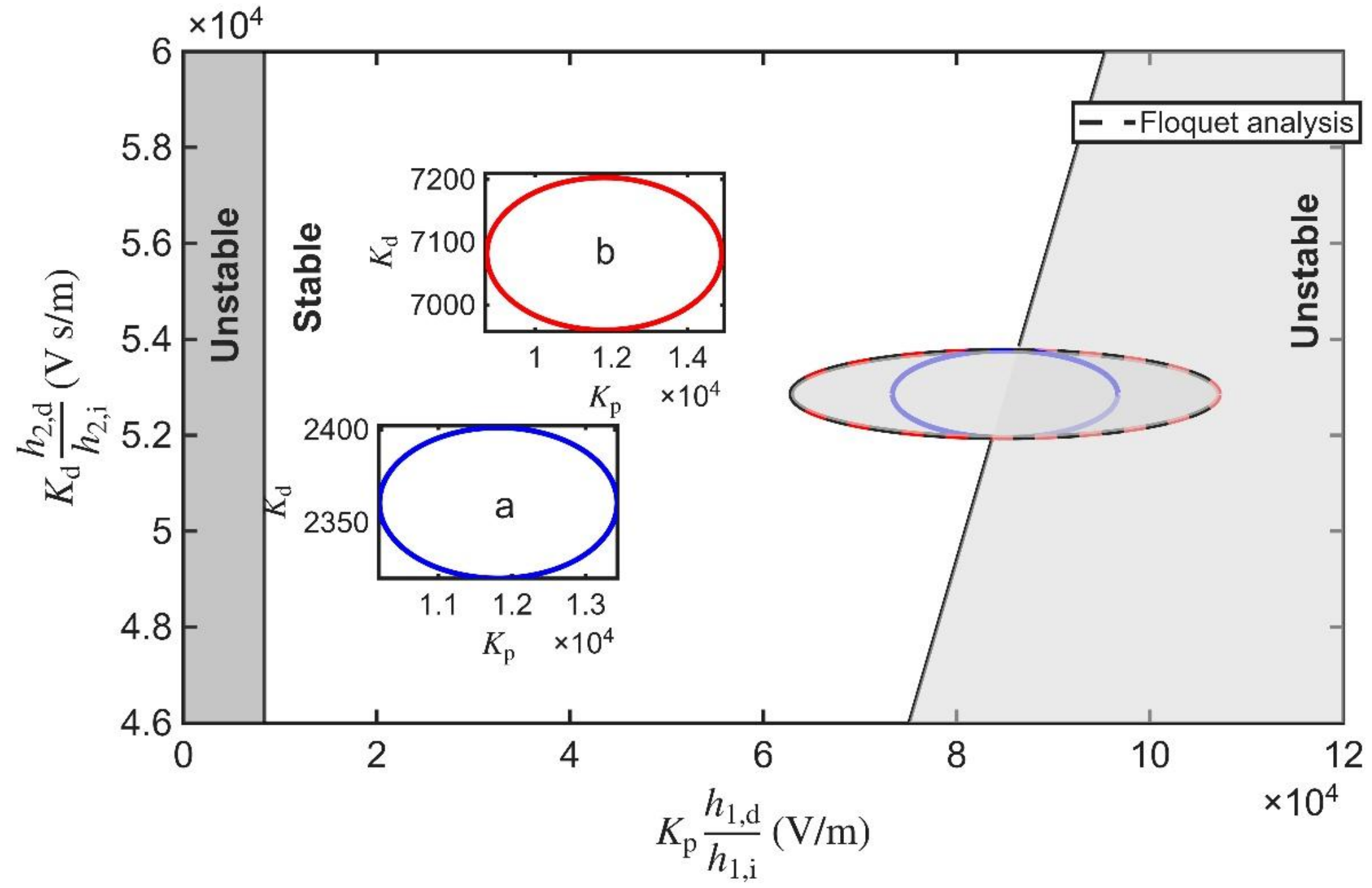

Figure 3. Simple parametric resonance stability boundaries (ellipses) observed for $\theta=0$. Combination parametric resonance is completely absent in this case. Here, $C=0.05\left(\mathrm{Nm}^2/\mathrm{A}^2\right)$, $z_0=0.015\left(\mathrm{m}\right)$, $m=7650\left(\mathrm{kg}\right)$, $R=9.71(\mathrm{Ohm})$, $\Omega=80\left(\mathrm{rad/s}\right)$, $A=0.0142\left(\mathrm{m}\right)$ and indices $i=a,b,c,d$ represent each ellipse. The factors $h_{1,\mathrm{i}}$ and $h_{2,\mathrm{i}}$, with $i=\mathrm{a,b,c,d}$, are used only to scale the $K_\mathrm{p}$ and $K_\mathrm{d}$ axes so that the different ellipses can be displayed together and compared clearly;

without this scaling, the ellipses would lie far apart in the parameter space. The dashed curve indicates the boundary obtained from numerical Floquet analysis.

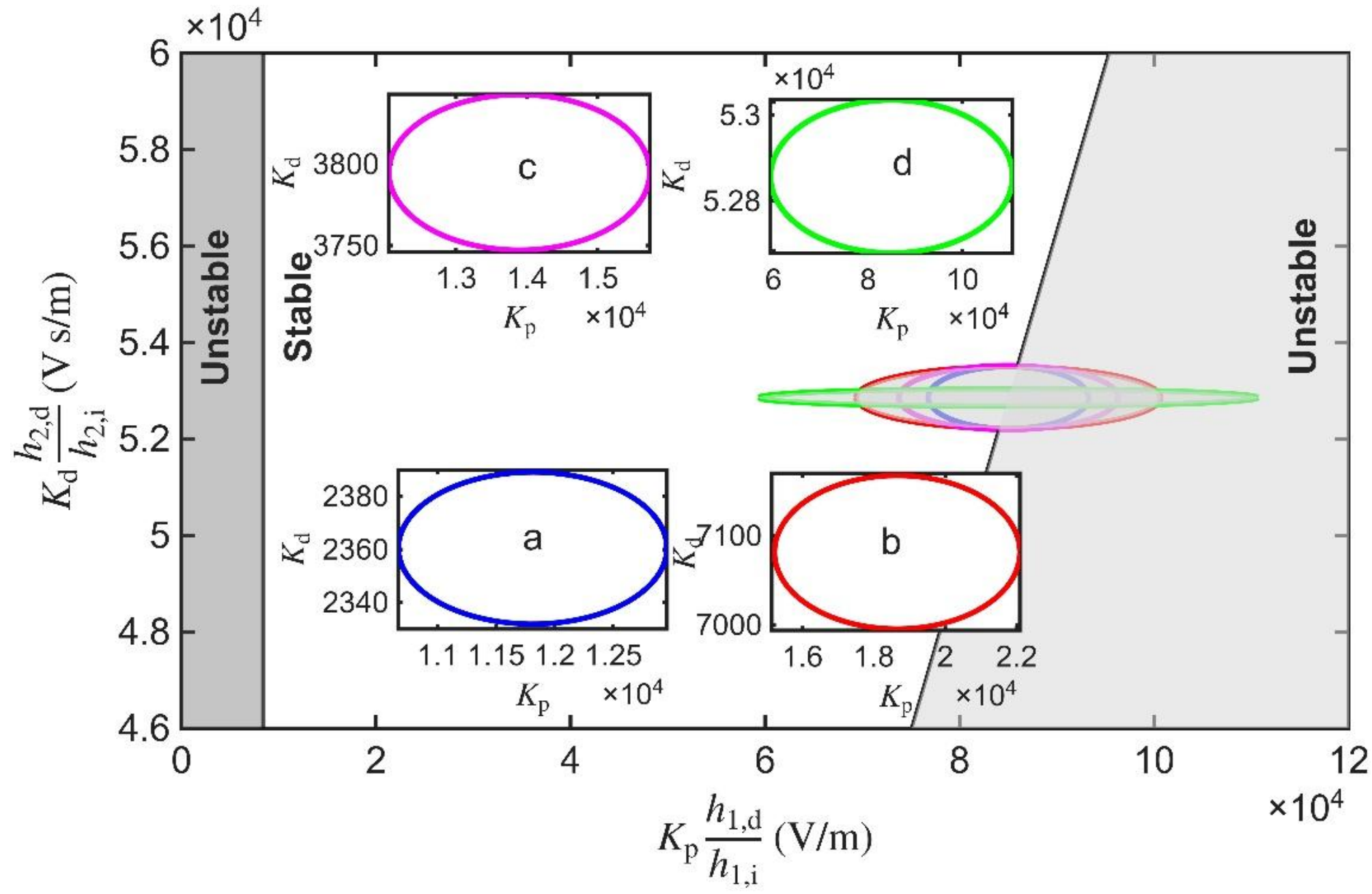

Figure 4. Simple and combination parametric resonance stability boundaries observed for $\theta = \pi / 2$. Here, $C = 0.05\left(\mathrm{Nm^2/A^2}\right)$, $z_0 = 0.015\left(\mathrm{m}\right)$, $m = 7650\left(\mathrm{kg}\right)$, $R = 9.71(\mathrm{Ohm})$, $\Omega = 80\left(\mathrm{rad/s}\right)$, $A = 0.0142\left(\mathrm{m}\right)$ and indices $i = \mathrm{a,b,c,d}$ represent each ellipse.

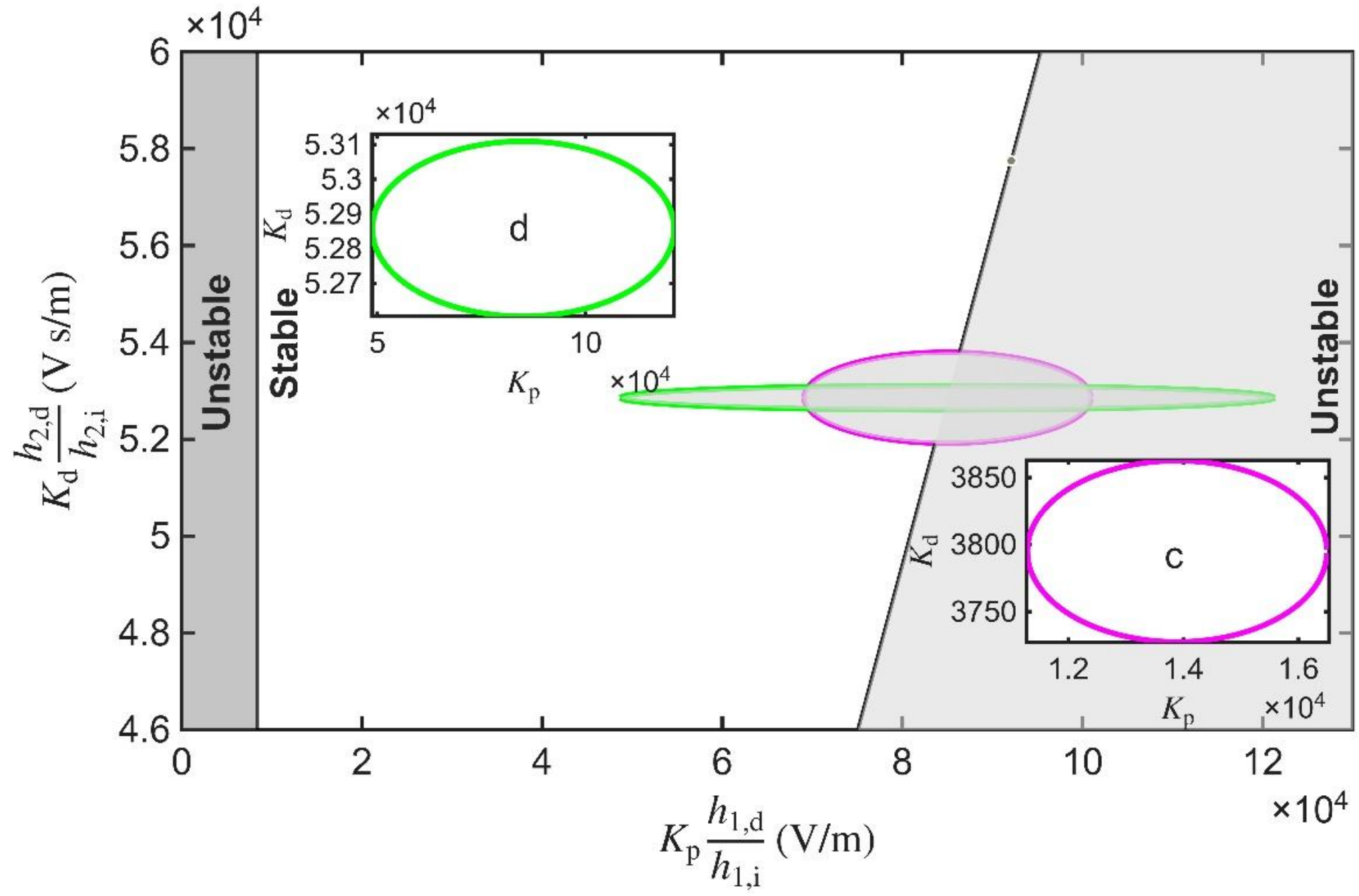

Figure 5. Combination parametric resonance stability boundaries observed for $\theta = \pi$. Simple parametric resonance is completely absent in this case. Here, $C = 0.05\left(\mathrm{Nm^2/A^2}\right)$, $z_0 = 0.015\left(\mathrm{m}\right)$, $m = 7650\left(\mathrm{kg}\right)$, $R = 9.71(\mathrm{Ohm})$, $\Omega = 80\left(\mathrm{rad/s}\right)$, $A = 0.0142\left(\mathrm{m}\right)$, and indices $i = \mathrm{a,b,c,d}$ represent each ellipse.

The simple parametric resonance ellipse is shown in Fig. 3 (ellipse a). Clearly, the ellipse indents the stable domain (of the time-invariant system) and renders it locally more narrow. The shaded regions in Figs. 3–5 indicate unstable domains. The left shaded region corresponds to divergence instability, while the right shaded region corresponds to dynamic instability. Numerical Floquet analysis [32] (explained at the end of Section 4.2 in more detail) confirms that only the (left) part of each ellipse bordering the stable domain acts as the effective stability boundary; beyond this boundary, a pair of Floquet exponents reaches modulus larger than one. This explains why the parts of the ellipses lying in the already unstable domain do not form an additional stability boundary.

From Eq. (20), we find that the coordinate $K_{\mathrm{p}} = h_{1,\mathrm{a}}$ has both a constant part $h_0 = mRg / \left(\sqrt{2}\sqrt{Cgm}\right)$ and a varying part depending on $K_{\mathrm{d}}$ through $\Omega$ [17]; note that $\Omega = 2\omega_2$. The $K_{\mathrm{d}}$ value of the centre of the ellipse (which depends on $\Omega$) can be found rom Eq. (20) too:

$$K_{\mathrm{d}} = h_{2,\mathrm{a}} \tag{21}$$

The size of the major axis of the ellipse relative to the local width of stable region (i.e., $h_0$), from the vertical line to the inclined line (see Fig. 3), can be compared for different values of excitation frequency $\Omega$ and thus locations along the inclined line in the $K_{\mathrm{p}} - K_{\mathrm{d}}$ plane. This relative size measure can be represented as:

$$\eta_{\mathrm{a}} = \frac{k_{1,\mathrm{a}}}{h_{1,\mathrm{a}} - h_0} = \frac{A\sqrt{1+\cos\theta}}{4\sqrt{2}z_0} \tag{22}$$

Eq. (22) offers the following key insights: (1) the relative size of the ellipse's major axis is frequency-independent and thus $v/L$ independent, even though the location of the ellipse changes with $\Omega$, (2) the relative size of the major axis decreases when increasing $\theta \in [0,\pi]$, with a maximum at $\theta = 0$ and minimum at $\theta = \pi$ (see Fig. 3), and (3) the relative size of the ellipse is linearly proportional to the normalized amplitude of the base oscillation $A/z_0$; clearly the ellipse spans in the worst case one quarter of the local width of the stable zone (i.e., when $A/z_0 = 1$ and $\theta = 0$).

To determine the second stability boundary (denoted as $\mathrm{b}$), related to the translational degree of freedom ($\omega_1 = \Omega/2$), we assume the following solution [19],

$$\begin{aligned} \Delta_{\mathrm{tr}}(t) &= a_0 \cos(\omega_1 t) + a_1 \sin(\omega_1 t) \\ \phi_{\mathrm{tr}}(t) &= 0 \end{aligned} \tag{23}$$

substitute Eq. (23) into Eq. (11), and follow a similar procedure, yielding the new expressions of $S_1$ and $C_1$ (see Eq. (16)):

$$\begin{aligned} &\left(-8K_{\mathrm{p}}L\sqrt{Cgm} + 4\sqrt{2}gLmR - \frac{ALmR\omega_1^2}{\sqrt{2}} - \frac{LmPR\omega_1^2}{\sqrt{2}} + 2\sqrt{2}LmRz0\omega_1^2\right)a_0 - \\ &\left(8K_{\mathrm{d}}L\sqrt{Cgm}\,\omega_1 + \frac{LmQR\omega_1^2}{\sqrt{2}} - 4\sqrt{2}CLm\omega_1^3\right)a_1 = 0 \end{aligned} \tag{24}$$

$$\left(8K_{\mathrm{d}}L\sqrt{Cgm}\omega_1-\frac{LmQR\omega_1^2}{\sqrt{2}}-4\sqrt{2}CLm\omega_1^3\right)a_0-$$
$$\left(8K_{\mathrm{p}}L\sqrt{Cgm}-4\sqrt{2}gLmR-\frac{ALmR\omega_1^2}{\sqrt{2}}-\frac{LmPR\omega_1^2}{\sqrt{2}}-2\sqrt{2}LmRz_0\omega_1^2\right)a_1=0 \tag{25}$$

This leads to another ellipse equation, and the properties of this ellipse follow similar relationships ( $\omega_1$ has been replaced by $\Omega/2$ ):

$$\frac{\left(K_{\mathrm{p}}-h_{1,\mathrm{b}}\right)^2}{k_{1,\mathrm{b}}^2}+\frac{\left(K_{\mathrm{d}}-h_{2,\mathrm{b}}\right)^2}{k_{2,\mathrm{b}}^2}=1 \tag{26}$$

$$\begin{aligned}
h_{1,\mathrm{b}}&=\frac{mR\left(8g+z_0\Omega^2\right)}{8\sqrt{2}\sqrt{Cgm}}\\
h_{2,\mathrm{b}}&=\frac{Cm\Omega^2}{4\sqrt{2}\sqrt{Cgm}}\\
k_{1,\mathrm{b}}&=\frac{\Omega}{2}k_{2,\mathrm{b}}\\
k_{2,\mathrm{b}}&=\sqrt{\frac{A^2mR^2\Omega^2\left(1+\cos\theta\right)}{256Cg}}
\end{aligned} \tag{27}$$

Here, the ellipse is larger than the first one related to Eq. (19) because $k_{1,\mathrm{b}}>k_{1,\mathrm{a}}$, but the ratio $\eta_{\mathrm{b}}$ is the same as for ellipse $\mathrm{a}$. Note that the second simple parametric resonance ellipse has the same location as the first one in Fig. 3 due to the applied scaling of the axes; the actual locations of the ellipses are different as shown in the inset plots (and determined by the corresponding resonance conditions $\Omega=2\omega_2$ and $\Omega=2\omega_1$ ).

### 4.2. Combination parametric resonance

In this section, the stability boundary related to combination parametric resonance is explored. Since it is a combination effect, we need to consider the coupled Eqns. (11)-(12) together. We assume the following of solution [19]:

$$\begin{aligned}
\Delta_{\mathrm{tr}}\left(t\right)&=a_0\cos\left(\omega_1t\right)+a_1\sin\left(\omega_1t\right)\\
\phi_{\mathrm{tr}}\left(t\right)&=b_0\cos\left(\omega_2t\right)+b_1\sin\left(\omega_2t\right)
\end{aligned} \tag{28}$$

For the combination resonance, we have the two possibilities as described in Section 4: $\omega_1+\omega_2=\Omega$, and $\omega_2-\omega_1=\Omega$. Using Eq. (14), we can find expressions for $\omega_1$ and $\omega_2$ which combine to $\Omega$ when added and subtracted, respectively:

$$\omega_1=\frac{1}{\sqrt{3}+1}\Omega;\ \omega_2=\frac{\sqrt{3}}{\sqrt{3}+1}\Omega; \tag{29}$$

$$\omega_1=\frac{1}{\sqrt{3}-1}\Omega;\ \omega_2=\frac{\sqrt{3}}{\sqrt{3}-1}\Omega; \tag{30}$$

The procedure to find the stability boundary is as follows; first, we substitute Eq. (28) in Eqns. (11)-(12) which gives two equations both having the following form:

$$S_1 \sin(\omega_1 t) + C_1 \cos(\omega_1 t) + S_2 \sin(\omega_2 t) + C_2 \cos(\omega_2 t) = 0 \tag{31}$$

Selecting $S_1$ and $C_1$ from Eq. (11), $S_2$ and $C_2$ from Eq. (12) (other terms are neglected after the trigonometric reduction, consistent with Eq. (31)), and substituting $\omega_2 = \sqrt{3}\omega_1$ gives the following four equations:

$$M_1 a_0 - M_2 a_1 + L_1 b_0 + L_2 b_1 = 0 \tag{32}$$

$$M_2 a_0 + M_1 a_1 + L_2 b_0 - L_1 b_1 = 0 \tag{33}$$

$$L_3 a_0 + L_4 a_1 + M_1 b_0 - M_3 b_1 = 0 \tag{34}$$

$$L_4 a_0 - L_3 a_1 + M_3 b_0 + M_1 b_1 = 0 \tag{35}$$

where

$$\begin{aligned} &M_1 = 2L\left(-4K_\mathrm{p}\sqrt{Cgm} + \sqrt{2}mR\left(2g + z_0\omega_1^2\right)\right);\ M_2 = 8K_\mathrm{d}L\sqrt{Cgm}\omega_1 - 4\sqrt{2}CLm\omega_1^3; \\ &M_3 = 8\sqrt{3}K_\mathrm{d}L\sqrt{Cgm}\omega_1 - 4\sqrt{6}CLm\omega_1^3;\ L_1 = \frac{L^2 m(-A+P)R\omega_1^2}{2\sqrt{2}}; \\ &L_2 = \frac{L^2 mQR\omega_1^2}{2\sqrt{2}};\ L_3 = \sqrt{2}m(-A+P)R\omega_1^2;\ L_4 = \sqrt{2}mQR\omega_1^2; \end{aligned} \tag{36}$$

The truncated Hill's determinant for the system given in Eq. (32)- (35) can be written as follows

$$\begin{vmatrix} M_1 & -M_2 & L_1 & L_2 \\ M_2 & M_1 & L_2 & -L_1 \\ L_3 & L_4 & M_1 & -M_3 \\ L_4 & -L_3 & M_3 & M_1 \end{vmatrix} = 0 \tag{37}$$

Even though Eq. (37) contains all information about the ellipses, we cannot extract the general form of the ellipse from this equation. To do so, we can use the following mathematical manipulation. Moving terms with $b_0$ and $b_1$ to the right-hand side, then squaring and adding Eqns. (32) and (33), and doing the same to Eqns. (34) and (35) yields

$$\left(M_1^2 + M_2^2\right)\left(a_0^2 + a_1^2\right) = \left(L_1^2 + L_2^2\right)\left(b_0^2 + b_1^2\right) \tag{38}$$

$$\left(L_3^2 + L_4^2\right)\left(a_0^2 + a_1^2\right) = \left(M_1^2 + M_3^2\right)\left(b_0^2 + b_1^2\right) \tag{39}$$

For non-trivial solutions, the determinant of Eqns. (38) and (39) should vanish:

$$\begin{vmatrix} M_1^2 + M_2^2 & -\left(L_1^2 + L_2^2\right) \\ L_3^2 + L_4^2 & -\left(M_1^2 + M_3^2\right) \end{vmatrix} = 0 \tag{40}$$

Using Eqns. (37) and Eq. (40), we can derive the following a mathematical condition which directly gives the ellipse equation (see Eq. (3.12) in [19]):

$$M_1^2 + M_2 M_3 = \sqrt{\left(L_1^2 + L_2^2\right)\left(L_3^2 + L_4^2\right)} \tag{41}$$

Substituting the first set of frequencies, given in Eq. (29), into Eq. (41) gives the first stability boundary (denoted as $\mathrm{c}$) related to combination parametric resonance:

$$\frac{\left(K_{\mathrm{p}}-h_{1,\mathrm{c}}\right)^2}{k_{1,\mathrm{c}}^2}+\frac{\left(K_{\mathrm{d}}-h_{2,\mathrm{c}}\right)^2}{k_{2,\mathrm{c}}^2}=1 \tag{42}$$

$$\begin{aligned}
h_{1,\mathrm{c}} &= \frac{mR\left(4g+\left(2-\sqrt{3}\right)z_0\Omega^2\right)}{4\sqrt{2}\sqrt{Cgm}} \\
h_{2,\mathrm{c}} &= \frac{\left(2-\sqrt{3}\right)Cm\Omega^2}{2\sqrt{2}\sqrt{Cgm}} \\
k_{1,\mathrm{c}} &= \sqrt{\frac{\sqrt{3}\left(2-\sqrt{3}\right)}{2}}\Omega k_{2,\mathrm{c}} \\
k_{2,\mathrm{c}} &= \sqrt{\frac{\left(2-\sqrt{3}\right)A^2mR^2\Omega^2\left(1-\cos\theta\right)}{128\sqrt{3}Cg}}
\end{aligned} \tag{43}$$

Substituting the set of frequencies given in Eq. (30) into Eq. (41) gives the second ellipse (denoted as $\mathrm{d}$):

$$\frac{\left(K_{\mathrm{p}}-h_{1,\mathrm{d}}\right)^2}{k_{1,\mathrm{d}}^2}+\frac{\left(K_{\mathrm{d}}-h_{2,\mathrm{d}}\right)^2}{k_{2,\mathrm{d}}^2}=1 \tag{44}$$

$$\begin{aligned}
h_{1,\mathrm{d}} &= \frac{mR\left(4g+\left(2+\sqrt{3}\right)z_0\Omega^2\right)}{4\sqrt{2}\sqrt{Cgm}} \\
h_{2,\mathrm{d}} &= \frac{\left(2+\sqrt{3}\right)Cm\Omega^2}{2\sqrt{2}\sqrt{Cgm}} \\
k_{1,\mathrm{d}} &= \sqrt{\frac{\sqrt{3}\left(2+\sqrt{3}\right)}{2}}\Omega k_{2,\mathrm{d}} \\
k_{2,\mathrm{d}} &= \sqrt{\frac{\left(2+\sqrt{3}\right)A^2mR^2\Omega^2\left(1-\cos\theta\right)}{128\sqrt{3}Cg}}
\end{aligned} \tag{45}$$

Similar to the cases of simple parametric resonance, for the combination parametric resonances the $K_{\mathrm{p}}$ and $K_{\mathrm{d}}$ values of the centre of the ellipses can be found easily too.

Furthermore, we have the following observations; contrary to the observation for simple resonance, the ellipses are now minimal at $\theta=0$ and maximum at $\theta=\pi$ (see Fig. 5). However, the relative sizes of the ellipses are also independent of the exciting frequency $\Omega$, like for the simple parametric resonance ellipses (Eq. (22)); the relative size measure is given by (note that $h_0$ is the local width of the stable region)

$$\eta_{\mathrm{c}}=\frac{k_{1,\mathrm{c}}}{h_{1,\mathrm{c}}-h_0}=\frac{A\sqrt{1-\cos\theta}}{4\sqrt{2}z_0}=\eta_{\mathrm{d}} \tag{46}$$

Note that the relative sizes of the ellipses are linearly proportional to the normalized amplitude of the base oscillation $A/z_0$; clearly the ellipses span again, in the worst case, one quarter of the local width of the stable zone (i.e., when $A/z_0 = 1$ and $\theta = \pi$).

We now have closed-form expressions for all the ellipses, allowing for direct comparison. Indices $\mathrm{a}$ and $\mathrm{b}$ represent ellipses related to simple parametric resonance, while $\mathrm{c}$ and $\mathrm{d}$ represent ellipses related to combination resonance. To compare the sizes, $\theta = \pi / 2$ is considered where all ellipses are present simultaneously. At this point, the order of ellipse sizes is as follows: $\mathrm{a} < \mathrm{c} < \mathrm{b} < \mathrm{d}$ (see inset plots in Fig. 4). The ratio of the major axis lengths of the two ellipses related to simple parametric resonance is $k_{1,\mathrm{b}} / k_{1,\mathrm{a}} = 3$, and for the combination parametric resonance ellipses, the ratio is $k_{1,\mathrm{d}} / k_{1,\mathrm{c}} \approx 14$; ellipse $\mathrm{d}$ is the largest ellipse.

All the results for the elliptic stability boundaries were verified numerically using standard Floquet analysis [32], and the analytical and numerical results were found to match exactly. For each selected pair of controller gains $\left(K_{\mathrm{p}}, K_{\mathrm{d}}\right)$, the linearized time-periodic system (Eqns. (8)-(10)) was integrated over one excitation period $T = 2\pi / \Omega$ using six independent sets of initial conditions to construct the monodromy matrix. The eigenvalues of this matrix, i.e., the Floquet multipliers, were then used to determine stability; the response is stable when all multipliers have modulus smaller than or equal to one and unstable when at least one multiplier has modulus larger than one. The contour $|\lambda| = 1$ obtained from the numerical Floquet analysis coincides with the analytical stability boundaries, as illustrated for one representative case in Fig. 3 (i.e., for parametric resonance specifically).

### 4.3. Practical impact

The relative size measures $\eta_{\mathrm{a}}$, $\eta_{\mathrm{b}}$, $\eta_{\mathrm{c}}$, and $\eta_{\mathrm{d}}$ (Eqs. (22) and (46))) indicate how strongly each parametric-resonance ellipse reduces the local width of the stable domain in the PD control parameter space. These values are independent of the excitation frequency and therefore of vehicle speed (for a fixed guideway wavelength), although the ellipses' locations and absolute sizes vary with speed.

Tables 1 and 2 give indicative values for practically relevant excitation amplitudes and phase shifts. The results show that combination parametric resonance can be less relevant than, as relevant as, or more relevant than simple parametric resonance, and may therefore impose an important design or operational constraint. The values are based on the simplified model considered here, with a rigid support and a single suspension layer, and should therefore be interpreted as first indications.

Table 1. Relative sizes of the parametric-resonance ellipses for different phase shifts at $A = 0.0142(\mathrm{m})$ and $z_0 = 0.015(\mathrm{m})$.

| $L / d$ | $\theta$ | $\eta_{\mathrm{a}} = \eta_{\mathrm{b}}$ | $\eta_{\mathrm{c}} = \eta_{\mathrm{d}}$ |
|---|---|---|---|
| 0 | 0 | 0.237 | 0 |
| 1/8 | $\pi / 4$ | 0.219 | 0.091 |
| 1/4 | $\pi / 2$ | 0.167 | 0.167 |
| 3/8 | $3\pi / 4$ | 0.091 | 0.219 |
| 1/2 | $\pi$ | 0 | 0.237 |

Table 2. Relative sizes of the parametric-resonance ellipses for different excitation amplitudes at $\theta = \pi / 4$ and $z_0 = 0.015(\mathrm{m})$.

| $A$ | $\eta_{\mathrm{a}}=\eta_{\mathrm{b}}$ | $\eta_{\mathrm{c}}=\eta_{\mathrm{d}}$ |
|---|---|---|
| 0.005 | 0.077 | 0.032 |
| 0.010 | 0.154 | 0.064 |
| 0.015 | 0.231 | 0.096 |

## 5. Suspensions with hybrid magnet: effect on stability boundary

In this section, the effect of using hybrid magnets—comprising the suspensions shown in Fig. 1 combined with permanent magnets—on the stability boundary is investigated [33–38]. Current Maglev and Hyperloop designs commonly use a combination of permanent and electromagnets. The permanent magnet can be designed to carry the static weight of the vehicle, while the electro-magnet takes care of perturbations around the static response. This way, the energy consumption is significantly reduced. This hybrid configuration raises the question of its influence on the system stability, which is the subject of investigation of this section. According to [2], the general form of the equations of motion for a hybrid-magnet system can be written as follows:

$$m\ddot{z}=-C\left(\left(\frac{I_1+\gamma}{\Delta_1+\beta}\right)^2+\left(\frac{I_2+\gamma}{\Delta_2+\beta}\right)^2\right)+mg \tag{47}$$

$$J\ddot{\phi}=-C\left(\left(\frac{I_1+\gamma}{\Delta_1+\beta}\right)^2-\left(\frac{I_2+\gamma}{\Delta_2+\beta}\right)^2\right)\frac{L}{2} \tag{48}$$

where $\beta$ and $\gamma$ are constants. The current equations for a single hybrid magnet can be written as:

$$\dot{I}_{1,2}+\frac{\Delta_{1,2}+\beta}{2C}\left(RI_{1,2}\right)=\frac{\Delta_{1,2}+\beta}{2C}U_{1,2}(t)+\left(\frac{I_{1,2}+\gamma}{\Delta_{1,2}+\beta}\right)\dot{\Delta}_{1,2} \tag{49}$$

where

$$U_{1,2}(t)=U_{\mathrm{ss1,ss2}}+K_{\mathrm{p}}(\Delta_{1,2}-\Delta_{\mathrm{ss1,ss2}})+K_{\mathrm{d}}(\dot{\Delta}_{1,2}-\dot{\Delta}_{\mathrm{ss1,ss2}}) \tag{50}$$

It is possible to rewrite Eq. (49) and (50) as

$$\dot{I}_{1,2}+\frac{\Delta_{1,2}+\beta}{2C}\left(RI_{1,2}+R\gamma\right)=\frac{\Delta_{1,2}+\beta}{2C}\left(U_{1,2}(t)+R\gamma\right)+\left(\frac{I_{1,2}+\gamma}{\Delta_{1,2}+\beta}\right)\dot{\Delta}_{1,2} \tag{51}$$

$$U_{1,2}(t)+R\gamma=U_{\mathrm{ss1,ss2}}+R\gamma+K_{\mathrm{p}}\left(\Delta_{1,2}+\beta-\left(\Delta_{\mathrm{ss1,ss2}}+\beta\right)\right)+K_{\mathrm{d}}(\dot{\Delta}_{1,2}-\dot{\Delta}_{\mathrm{ss1,ss2}}) \tag{52}$$

Assuming the following variable transformations

$$\bar{\Delta}_{1,2}=\Delta_{1,2}+\beta;\ \dot{\bar{\Delta}}_{1,2}=\dot{\Delta}_{1,2};\ \bar{I}_{1,2}=I_{1,2}+\gamma;\ \dot{\bar{I}}_{1,2}=\dot{I}_{1,2};\ \bar{U}_{1,2}=U_{1,2}+R\gamma \tag{53}$$

Eqns. (47)-(50) become

$$m\ddot{z}=-C\left(\left(\frac{\bar{I}_1}{\bar{\Delta}_1}\right)^2+\left(\frac{\bar{I}_2}{\bar{\Delta}_2}\right)^2\right)+mg \tag{54}$$

$$J\ddot{\phi} = -C\left(\left(\frac{\bar{I}_1}{\bar{\Delta}_1}\right)^2 - \left(\frac{\bar{I}_2}{\bar{\Delta}_2}\right)^2\right)\frac{L}{2} \tag{55}$$

$$\dot{\bar{I}}_{1,2} + \frac{\bar{\Delta}_{1,2}}{2C}\left(R\bar{I}_{1,2}\right) = \frac{\bar{\Delta}_{1,2}}{2C}\bar{U}_{1,2}(t) + \left(\frac{\bar{I}_{1,2}}{\bar{\Delta}_{1,2}}\right)\dot{\bar{\Delta}}_{1,2} \tag{56}$$

$$\bar{U}_{1,2}(t) = \bar{U}_{\text{ss1,ss2}} + K_{\text{p}}(\bar{\Delta}_{1,2} - \bar{\Delta}_{\text{ss1,ss2}}) + K_{\text{d}}(\dot{\bar{\Delta}}_{1,2} - \dot{\bar{\Delta}}_{\text{ss1,ss2}}) \tag{57}$$

The steady-state gap for the system with the hybrid magnets is as follows: $\Delta_{\text{ss1,ss2}} = \bar{\Delta}_{\text{ss1,ss2}} - \beta$.

The system of Eqns. (54)-(57) has exactly the same structure as the system of Eqns. (1), (2), (4), and (5). The linearized, simplified EOMs (Eqns. (11) and (12)) are also very similar. However, it should be clear that the time-dependent coefficients are based on the modified steady states; in other words, the following replacement should be made to get the correct coefficients: $\Delta_{\text{ss1,ss2}} \to \bar{\Delta}_{\text{ss1,ss2}}$. In essence, this comes down to the following change: $z_0 \to z_0 + \beta$.

The effects of the slightly different coefficients on the stable domain are as follows (verified by numerical Floquet analysis):

- the left boundary is not affected;
- the right, straight boundary rotates to the right, which is reasonable as its slope is inversely proportional to $z_0$ [17] and implies an overall stabilizing effect of the hybrid magnets;
- the only change in the equations for the parametric-resonance ellipses (Eqns. (19), (26), (42) and (44)) is $z_0 \to z_0 + \beta$;
- the absolute magnitude of the ellipses does not change as $z_0$ only affects the $K_{\text{p}}$ coordinate of the ellipses; i.e., $h_{1,i}$ $(i = \text{a,b,c,d})$;
- the ellipses therefore only shift to the right, in $K_{\text{p}}$ direction, in accordance with the rotation of the right boundary and such that the centers stay on that boundary.

While the sizes of the ellipses do not change, their relative sizes do, as the local widths of the stable zone increase due to the rotation of the right boundary. Introducing the replacement $z_0 \to z_0 + \beta$ into the expressions for $\eta_i$ $(i = \text{a,b,c,d})$ (Eqns. (22) and (46)), we find that relative sizes decrease when a hybrid system is used instead of a fully electromagnetic system. For example, the $\beta = z_0/2$ (which is close to the value employed by [33]), the relative sizes decrease to $2/3$ of their original values, while for $\beta = z_0/4$, the relative sizes decrease to $4/5$ of their original values.

Note that the value of $\gamma$ that guarantees $I_1$ and $I_2$ to be zero (i.e., the permanent magnets carry the static load) in the specific steady state without base excitation is given by:

$$\gamma = \sqrt{\frac{mg}{2C}}\left(z_0 + \beta\right) \tag{58}$$

## 6. Conclusion

This study provides a comprehensive analysis of the parametric resonances of an electromagnetically suspended vehicle subjected to periodic base excitations. By modelling the vehicle dynamics with translational and rotational degrees of freedom, we capture both simple and combination parametric resonance phenomena. Our analysis shows that oscillations in the supporting structure can lead to instabilities depending on the alignment of natural frequencies and excitation frequencies. Simple resonance occurs when either translational or rotational modes are independently excited, while combination parametric resonance emerges when both modes interact. We have derived analytical expressions for the stability boundaries using an extended Hill's method, and presented them in terms of system parameters such as frequency and amplitude of and phase shift between the base excitations; it turns out that the parametric-resonance stability boundaries are ellipses in the PD control parameter space that indent the otherwise triangular stable zone. There is essentially no speed dependence in the relative size of the parametric-resonance ellipses (normalized by the local width of the stable domain), and the combination parametric resonance ellipses can be smaller than, as large as, or larger than the simple parametric resonance ellipses depending on the phase shift. However, the absolute size of the ellipses increases significantly with speed. Comparing the four ellipses obtained from simple and combination parametric resonances, one of the ellipses corresponding to combination parametric resonance is the largest. Additionally, we show that incorporating hybrid magnets in the vehicle suspensions does not affect the size of the parametric-resonance zones; it only shifts them in accordance with an overall widening of the stable domain. This highlights the feasibility of the hybrid magnet for energy-efficient suspension designs.

The findings of this paper underline the importance of carefully selecting design and operating conditions to avoid instability, especially in systems like Hyperloop/Maglev where small support deviations can pose significant (parametric) excitation. The present work should be viewed as an initial study into the relevance of simple and, in particular, of combination parametric resonances under sinusoidal base excitations. The results suggest that avoiding combination parametric resonance particularly may become an important design and operational constraint for Maglev and Hyperloop systems. To obtain more quantitative design recommendations, future work should include guideway dynamics, aeroelastic loading, non-harmonic and random base excitation, and additional suspension layers.

## 7. Funding declaration

The authors express sincere gratitude to the European Union's Horizon Europe programme for its support through the Marie Skłodowska-Curie grant agreement No 101106482 (HySpeed project).